\documentclass{article}
\usepackage{epsfig}
\usepackage{amssymb}
\usepackage{fullpage}

\begin{document}

%\begin{frontmatter}

\title{Extended hydrodynamics  from Enskog's equation \\ 
The bidimensional case\thanks{\fbox{published in Physica A, \underline{354},
77-87, 2005}}}
\author{Hideaki Ugawa \\
Departamento de F\'{\i}sica, FCFM, Universidad de
Chile, Santiago, Chile}

\maketitle

\begin{abstract}

A heat conduction problem is studied using extended hydrodynamic equations
obtained from Enskog's equation for a simple case of two planar systems in
contact through a porous wall. One of the systems is in equilibrium and the
other one in a steady conductive state. The example is used to put to test
the predictions which has been made with a new thermodynamic formalism. 
\end{abstract}


kinetic theory; Enskog dense gases; Heat conduction; Nonequilibrium thermodynamics

\section{Introduction}

Today Enskog's original kinetic theory is known as the standard
Enskog theory (SET)~\cite{chapman,ferziger,resibois} because after the
pioneer work of van Beijeren and Ernst~\cite{VB} there are several new
versions of Enskog's theory collectively called revised Enskog's theory
(RET)~\cite{bellomo}.  Among the latter there are versions that have been
extended to describe condensed matter~\cite{kirkpatrick}. To
Navier-Stokes level both SET and RET lead to the same
results~\cite{VB,garzo}, whether or not an external force is present. 
%\footnote{*Corresponding author.
%
%{\em Email address}: hugawa@cec.uchile.cl}

In the present article we make use of {\em extended hydrodynamic equations}
for the bidimensional case~\cite{hu}. They are more complete than a linear
approximation but still they are the result of an approximation scheme that
we explain elsewhere. Using a strategy as in Ref.~\cite{RC} and
approximations defined in Ref.~\cite{hu} we obtain in Sec.
\ref{sec:consideration} the same hydrodynamic equations for SET and RET.

In this article we apply our extended hydrodynamics to a one dimensional
steady heat conductive state.  There is much work on this as, 
for example, the
experimental results in~\cite{teagan} or the theoretical ones
in~\cite{gross,ohwada1,ohwada2}. Recently Kim and Hayakawa~\cite{kim} 
studied this
problem for hard core and Maxwellian particles using Boltzmann's equation
combined with Chapman-Enskog's method.  They tried a test and criticized the
analysis of the nonequilibrium steady state thermodynamics (SST) proposed by
Sasa and Tasaki~\cite{sasa}.  In the last reference the authors state that
if there is gas in a one dimensional heat conductive configuration in
contact, through a porous wall, with an equilibrium gas state, then  a
pressure difference  must appear in the direction of the heat flow.  We
analyze this double system making use of the extended hydrodynamic equations
derived from Enskog's equation 
using Grad's moment expansion method~\cite{grad}. 
Our conclusions  differ from those in~\cite{sasa}.

The organization of the present article is as follows.  
In Sec. \ref{sec:threep} the configuration of these systems 
is drawn schematically, 
in Sec. \ref{condition} the condition for the two systems 
to be in contact via the central porous plate is introduced: 
the upper and lower plates are normal plates; 
the central plate has many small pores through which the gas can pass. 
In Sec. \ref{sec:BEQs} we give the basic equations used in this paper.
Comments are in Sec.  \ref{sec:consideration}.  
Finally, our discussion and conclusions are 
written in Sec. \ref{sec:discussion} and \ref{sec:conclusion}, respectively.

\section{Definition of the system}\label{sec:threep}

\begin{figure}[htb]
\begin{center}
\epsfig{file=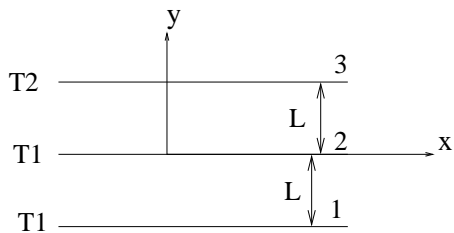,width=6.0cm,angle=0}

\caption{The plates of the system as described in the text. \label{fig:1}}
\end{center}
\end{figure}

Sasa and Tasaki~\cite{sasa} proposed an interesting system
consisting of a nonequilibrium steady state subsystem in contact with a
subsystem in equilibrium as explained below. This system has three plates as
shown in Fig.~\ref{fig:1} and there is gas between them.  The upper and lower
plates (plates 1 and 3) are normal plates.  The central plate (plate 2) has
pores through which gas can pass. 

Following Sasa and Tasaki, we consider the system consisting  of three
infinite parallel plates 1, 2 and 3 separated by a distance $L$. The $Y$
axis is defined perpendicular to them while an $X$ axis is placed on plate
2. The pores in plate 2 are distributed homogeneously. Plates 1 and 2 have
fixed temperature $T_{1}$ while plate 3 has a different (fixed) temperature
$T_2$.

After a sufficiently long time, by effusion, some of the gas passes through
plate 2 and the gas between plates 1 and 2 reaches an equilibrium state. 
The system between the plates 2 and 3 reaches a nonequilibrium
steady state with translation symmetry along the $X$ axis.

We assume that the typical distances between pores is very small and that
the diameter of the pores is also very small, so that the ratio between such
lengths and the mean free path is much smaller than unity. Having no
external force and no hydrodynamic velocity there is no heat flux parallel
to the plates. The system is in a static configuration.

%%%%%%%%%%%%%%%%%%%%%%%%%%%%%%%%%%%%%%%%%%%%%%%%%%%%%%%%%%%%%%%%%%%
%%%%%%%%%%%%%%%%%%%%%%%%%%%%%%%%%%%%%%%%%%%%%%%%%%%%%%%%%%%%%%%%%%%
\section{The contact condition}  \label{condition}

In general, there is a difference between the temperatures of the plates and
the temperatures of the gas in contact with them, this is a well-known
effect called thermal slip. However, for simplicity sake, we assume that
the temperature of plate 2 and the gas in contact on both sides of it are
equal, namely we are neglecting the Knudsen layer.
 
The velocity and the peculiar velocity of the gas will be denoted 
by $\mathbf{c}$ and $\mathbf{C}$, respectively.
The condition that there is no mass flux through plate 2 is  
\begin{equation}\label{tsuri}
\int_{-\infty}^{\infty}dc_{x}\int_{0}^{\infty}dc_{y}c_{y}f_{\rm equil.} +
\int_{-\infty}^{\infty}dc_{x}\int_{-\infty}^{0}dc_{y}c_{y}f_{y=0} = 0, 
\end{equation} 
where 
$f_{\rm equil.}=n_{\sf  eq}\left[\frac{1}{2\pi T_{1}}\right]
                      \exp{\left[-\frac{C^{2}}{2T_{1}}\right]}$ 
is the equilibrium distribution function associated to the gas 
between plates 1 and 2 and $n_{\sf  eq}$ is the uniform number density
in this same region  while $f_{y=0}$ is the nonequilibrium 
distribution function  between plates 2 and 3 
evaluated at $y=0$. 
Next, it is necessary to see how the two distributions satisfy
condition~(\ref{tsuri}).

Imposing condition (\ref{tsuri}) yields
\begin{equation}\label{keyos} 
n_{\sf  eq} = \frac{1}{2}\left[n(0) + P_{yy}\right] 
             -\delta \frac{\chi}{2\rm K\!n}\left[n(0) + P_{yy}\right]n(0) 
             +\delta^2\frac{\chi^{2}}{2\rm K\!n^{2}}\left[ n(0) +
P_{yy}\right]n(0)^{2}.
\end{equation}
In addition, the total mass conservation law for the system is
\begin{equation}\label{masssum}
   n_{\sf  eq}+\int_{0}^{1}n(y)dy=2.
\end{equation}

Above we are using dimensionless fields and dimensionless variables in
general. The fields $n$ (number density), $P_{ij}$ (pressure tensor), 
 $\vec Q$ (net heat flux vector), and 
$T$ (temperature) generally depend
on the coordinate $y$, where $p_{ij}$ and $\mathbf{q}^{k}$ are 
the symmetric and traceless part of the pressure 
tensor  and the kinetic part of the heat flux vector respectively.
These hydrodynamic fields are defined according to the following sum rules:
\begin{eqnarray}
\int fd\mathbf{c} &=& n(y)\,, \\
\int \mathbf{C}(y)fd\mathbf{c} &=& 0\,, \\
\int \frac{1}{2}C(y)^{2}fd\mathbf{c} 
         &=& n(y)T(y)\,,\\
\int C_{i}(y)C_{j}(y)fd\mathbf{c} 
 &=& n(y)T(y)\delta_{ij}+p_{ij}(y)\,, \\
\int \frac{1}{2}C(y)C^{2}(y)f
d\mathbf{c} &=& q_{y}^{\rm k}(y).
\end{eqnarray}

We also use the following dimensionless numbers
$$
{\rm K\!n} = \frac{8\sqrt{2}}{\pi}\frac{\ell}{L}\,,\quad
  \textrm{Knudsen number},\qquad
  \delta = \frac{\sigma}{L} = {\rm K\!n}\rho_{0} 
$$ 
where $\sigma$ is the particle's diameter, $\ell$ their mean free at
equilibrium and $\rho_0$ is the mean area density.

\section{Balance equations}\label{sec:BEQs}

The basic concrete equations solved here are the following.

 $\bullet$  In the case of the linearized Boltzmann-Grad method (LBG):
$P_{yy}(y) \equiv P_{yy} = \hbox{constant}$, 
$P_{xy}(y) \equiv P_{xy} = \hbox{constant}$,
$Q_{y}(y)  \equiv Q_{y} = \hbox{constant}$,
\begin{equation}\label{LBGeq}
n(y)T(y) =P_{xx}=P_{yy}, \qquad
\qquad
\frac{dT(y)}{dy}+\frac{2Q_{y}}{\rm K\!n \sqrt{\pi T(y)}}=0.
\end{equation}

 $\bullet$ In the case of the Enskog-Grad method (EG):
$P_{yy}(y)\equiv P_{yy}=\hbox{constant}$, \quad $Q_{y}(y)\equiv Q_{y}=\hbox{constant}$,
$P_{xy}(y)=p_{xy}(y)=0$,
\begin{eqnarray} 
 P_{yy}&=&-\left[1+\frac{\delta}{\rm K\!n}\chi n(y) \right] 
                          {p_{xx}} \left( y \right) 
+ \left[1+ 2\frac{\delta}{\rm K\!n}\chi n(y)\right]
           n \left( y \right) T \left( y \right), \label{EGeq1}  \\
Q_{y}&=& \left[1+\frac{3}{2}\chi n(y)\frac{\delta}{\rm K\!n}\right]q_{y}^{k}(y)
      -\delta^{2}\frac{2}{\sqrt{\pi}\rm K\!n}\chi 
                n(y)^{2}\sqrt{T(y)}\frac{dT(y)}{dy}, \label{EGeq2}  \\
-\frac{1}{2}\frac{dq_{y}^{k}(y)}{dy}
 &=& 
-\frac{8}{\sqrt{\pi}\rm K\!n}\chi \sqrt { T \left( y \right) }
          \left[n \left( y \right)  p_{xx} \left( y \right)
                -\frac {q_{y}^{k}(y)^{2}}{128  T\left( y\right)^{2}}\right]
\nonumber \\
 &&+\frac{\delta}{4\rm K\!n}\chi \left[5q_{y}^{k}(y)\frac{dn(y)}{dy}
 +3n(y)\frac{dq_{y}^{k}(y)}{dy} \right]  \label{EGeq3}
 \nonumber \\
 &&- \frac{\delta^{2}}{\sqrt{\pi}\rm K\!n}\chi n(y)\sqrt{T(y)}
       \left[ 2\frac{dn(y)}{dy}\frac{dT(y)}{dy}
             +\frac{1}{2}\frac{n(y)}{T(y)}\left(\frac{dn(y)}{dy}\right)^{2} 
             +n(y)\frac{d^{2}T(y)}{dy}\right],\nonumber \\  
\end{eqnarray}%%eqa
and
\begin{eqnarray} \label{EGeq4}
 &&-T(y)\frac{dp_{xx}(y)}{dy} -p_{xx}(y)\frac{dT(y)}{dy}
  -n(y)T(y)\frac{dT(y)}{dy}+\frac{T(y)}{n(y)}p_{xx}(y)\frac{dn(y)}{dy}
\nonumber \\
 && -\frac{p_{xx}(y)}{n(y)}\frac{dp_{xx}(y)}{dy}
\nonumber \\
&=&-\frac{4}{\sqrt{\pi T(y)}\rm K\!n}\left[ n(y)T(y)-p_{xx}(y) \right]
q_{y}^{k}(y) 
\nonumber \\
&&+ \frac{\delta}{\rm K\!n}\chi 
\left[  \frac{7}{2}n(y) T \left( y \right)\frac{dp_{xx}(y)}{dy} 
       +\frac{7}{2}T \left( y \right)p_{xx}(y)\frac{dn(y)}{dy}
       +2 p_{xx} \left( y \right)\frac{dT(y)}{dy} \right. \nonumber \\
       &&\left.-8n(y)T(y)^{2}\frac{dn(y)}{dy}
       -7n \left( y \right)^{2}T(y)\frac{dT(y)}{dy}\right].  
\end{eqnarray}%%eqa
The substitution of $\delta=0$ and $\chi=1$ in the EG equations 
yields the equations corresponding to the nonlinearized Boltzmann-Grad 
method (NLBG).

%%%%%%%%%%%%%%%%%%%%%%%%%%%%%%%%%%%%%%%%%%%%%%%%%%%%%%%%%%%%%%%%%%%%
%%%\section{Results}\label{sec:results}

%%%%%%%%%%%%%%%%%%%%%%%%%%%%%%%%%%%%%%%%%%%%%%%%%%%%%%%%%%%%%%%

\section{The pressure difference 
between the equilibrium and nonequilibrium sides}\label{sec:consideration}

All the results we describe in what follows were obtained using  perturbation
methods choosing $T_2>T_1$ and using $\epsilon= (T_{2}-T_{1})/T_{1}$ as
the perturbation parameter. We solve the system of equations and their boundary
(contact) conditions up to $\epsilon^{6}$. We choose $\delta=0.001$. In such
case $\rm K\!n$ is in inverse proportion to $\rho_{0}$.  We choose 
Henderson's expression~\cite{henderson} as the concrete expression for
$\chi$:
\begin{equation}\label{chi}
\chi = \frac{1-\frac{7}{16}\rho_{0}}{(1-\rho_{0})^{2}}.
\end{equation}

We calculate the pressure in both sides of plate 2.
Using $P_{yy}$ for the nonquilibrium steady state side and 
the pressure $P_{\sf eq}$ which is estimated by the state equation
for the equilibrium side
\begin{equation}\label{peq}
P_{\sf eq}\equiv \left[1+\delta\frac{2}{\rm K\!n}\chi n_{\sf  eq}\right]n_{\sf  eq},
\end{equation}
the  pressure difference $\Delta P$ is defined by 
\begin{equation}\label{osmop}
\Delta P=P_{yy}-P_{\sf eq}\,.
\end{equation}
Note that $n(0)=n_{\sf  eq}$ and $\Delta P=0$ to 
first order in $\epsilon$.
Hence, we rewrite our results in the following way
\begin{equation}\label{lambdan}
n(0)=n_{\sf  eq}\left[ 1 + \lambda_{n}\frac{Q_{y}^{2}}{n_{\sf  eq}^{2}}\right],
\end{equation}
\begin{equation}\label{lambdadp}
\Delta P=\lambda_{\Delta P}\frac{Q_{y}^{2}}{n_{\sf  eq}}
\end{equation}
where $\lambda_{n}$ and $\lambda_{\Delta P}$ are constants.
Furthermore, it is possible to rewrite $P_{yy}$:
\begin{equation} \label{lambdapyy}
P_{yy}=n(0)\left[1+\delta\frac{2}{\rm K\!n}\chi n(0)\right]
           \left[1+\lambda_{p}^{yy}\frac{Q_{y}^{2}}{n(0)^2}\right].
\end{equation}

Tables \ref{Table 1.} and \ref{Table 2.} give the values of these constants 
for $\epsilon=0.05$ and 0.1, respectively. 
 The value and sign of $\Delta P$ depend on
$\epsilon$ and ${\rm K\!n}$. 
Table \ref{Table 3.} gives the value of $\lambda_{\Delta P}$ obtained 
by first order EG, 
that is, up to $\delta$ for $\epsilon=0.05$ and 0.1, respectively. 
In this case, the pressure difference 
also exists and its value and sign depend on
$\epsilon$ and ${\rm K\!n}$, too.

\begin{table}[ht]
\begin{center}
\begin{tabular}{c||c|c|c}
$\rm K\!n$ & $\lambda_{n}\times 10^{-2}$ & $\lambda_{\Delta P}\times 10^{-2}$ & $\lambda_{p}^{yy}/\lambda_{n}$ \\
\hline \hline
0.005 & -0.360727 & 1.43059 & -3.45019 \\ 
\hline
0.01  & 0.6175405 &-0.37938 & -1.17319 \\
\hline
0.02  & 0.5891347 &-0.51577 & -1.66426 \\
\hline
0.05  & 0.4851223 &-0.46437 & -1.87508 \\
\hline
0.1   & 0.4309258 &-0.44057 & -1.93856 \\
\hline
0.2   & 0.4158376 &-0.41159 & -1.96948 \\
\hline
\end{tabular}
\end{center}
\caption{The values of 
$\lambda_{n}$, $\lambda_{\Delta P}$ and $\lambda_{p}^{yy}/\lambda_{n}$ 
for $\epsilon=0.05$ in the case of EG }
\label{Table 1.}
\end{table}

\begin{table}[ht]
\begin{center}
\begin{tabular}{c||c|c|c}
$\rm K\!n$ & $\lambda_{n}\times 10^{-2}$ &$\lambda_{\Delta P}\times 10^{-2}$ & 
$\lambda_{p}^{yy}/\lambda_{n}$ \\
\hline \hline
0.005 &-0.405391 & 1.49646 & -3.19368 \\ 
\hline
0.01  & 0.614064 &-0.37163 & -1.16022 \\
\hline
0.02  & 0.590175 &-0.51564 & -1.66031 \\
\hline
0.05  & 0.486046 &-0.40465 & -1.87365 \\
\hline
0.1   & 0.431343 &-0.43093 & -1.93785 \\
\hline
0.2   & 0.416123 &-0.41182 & -1.96124 \\
\hline
\end{tabular}
\end{center}
\caption{The values of 
$\lambda_{n}$, $\lambda_{\Delta P}$ and $\lambda_{p}^{yy}/\lambda_{n}$ 
for $\epsilon=0.1$ in the case of EG}
\label{Table 2.}
\end{table}

\begin{table}[ht]
\begin{center}
\begin{tabular}{c||c|c}
$\rm K\!n$ & $\lambda_{\Delta P}\times 10^{-2}$ $\epsilon=0.05$ 
           & $\lambda_{\Delta P}\times 10^{-2}$ $\epsilon=0.1$ \\ 
\hline \hline
0.005 & 1.208777 & 1.269384 \\ 
\hline
0.01  &-0.417305 &-0.410453 \\
\hline
0.1   &-0.431207 &-0.431631 \\
\hline
\end{tabular}
\end{center}
\caption{The values of 
$\lambda_{\Delta P}$ for $\epsilon=0.05$ and 0.1 in the case of the first order EG}
\label{Table 3.}
\end{table}

In the case of LBG, since $p_{xx}=0$ then  $\lambda_{n}=\lambda_{\Delta
P}=\lambda_{p}^{yy}=0$. There is no pressure difference in this case.

On the other hand, for the case of NLBG, the substitution of $\delta=0$ in Eqs.
(\ref{keyos}), (\ref{EGeq1}), (\ref{EGeq2}), (\ref{EGeq3}), (\ref{peq}), 
(\ref{osmop}), (\ref{lambdan}), (\ref{lambdadp}) and
(\ref{lambdapyy}) leads to $\lambda_{n}=1/256$, $\lambda_{\Delta P}=-1/256<0$
and $\lambda_{p}^{yy}=-1/128$ 
where for $\lambda_{n}$ and $\lambda_{\Delta P}$ it is correct to
consider only up to
second order in $Q_{y}$. Hence the osmotic pressure difference 
does exist and its value is constant and negative.

Furthermore, for the case of EG, $\lambda_{p}^{yy}/\lambda_{n}=
-2-\delta\frac{4}{\rm K\!n}n_{\sf  eq}\neq -2$.

We analyze the pressure difference $\Delta P$ from another point of view.
It is sufficient to calculate $\Delta P$ up to $\epsilon^2$.
It is given by
\begin{eqnarray}\label{2ndoe}
\Delta P &=& \epsilon^{2}\frac{\pi \rm K\!n^2}{4096}
             \left[215\rho_{0}^{2}-\frac{52}{\chi}\rho_{0}
                                  -\frac{1}{\chi^{2}}\right]
\\
         &=& \epsilon^{2}\frac{\pi \rm K\!n^2}{4096(7\rho_{0}-16)^{2}}
[15335\rho_{0}^{4}-69024\rho_{0}^{3}+81344\rho_{0}^{2}
-9216\rho_{0}-1024]. \nonumber 
\end{eqnarray}
It is seen that the sign of $\Delta P$ changes from  negative to 
positive  approximately at $\rho_{0}= 0.2$, whereas 
it is always negative in the NLBG and to first order in the EG's case.

Besides the system far from equilibrium, we are also interested 
in a region extremely close to the equilibrium condition.
Therefore, we analyze the case without the strong nonlinear term, namely, 
we eliminate the terms involving $q_{y}^{k}(y)^{2}$ 
and $p_{xx}(y)q_{y}^{k}(y)$ in Eqs. (\ref{EGeq3}) and (\ref{EGeq4}).    
In this case, up to $\delta$,
\begin{eqnarray}\label{leg1}
P_{yy}&=&  n(0)
        -\delta\frac{\sqrt{\pi}Q_{y}}{16n(0)}
                         \left(\frac{dn(y)}{dy}\right)_{y=0},
\nonumber \\
\Delta P &=& -\delta \frac{\sqrt{\pi}Q_{y}}{32n(0)}
                             \left(\frac{dn(y)}{dy}\right)_{y=0}.
\end{eqnarray} 
Assuming that the derivative $dn(y)/dy$ of the density at plate 2 has the same 
sign as $Q_{y}$ (this is normaly correct),  $\Delta P$ is always negative.
Evaluating up to  $\delta^2$ and $\epsilon^2$ yields
\begin{eqnarray}\label{leg2}
\Delta P&=& \epsilon^2\delta\frac{\pi \rm K\!n}{128\chi}\left[9\chi \rho_{0}-2\right]
      = \epsilon^2\delta \frac{\pi \rm K\!n}{128\chi}
         \left[  \frac{9\rho_{0}(16-7\rho_{0})}{16(1-\rho_{0})^{2}} -2 \right],
\nonumber \\
\lambda_{n}&=& \frac{\chi}{16}\left[1-\frac{9}{2}\chi \rho_{0}\right],
\qquad
\frac{\lambda_{p}^{yy}}{\lambda_{n}} = -2+2\chi \rho_{0}+9(\chi \rho_{0})^{2}.
\end{eqnarray} 
As $0<\rho_{0}<1$, $\Delta P$ is positive  but $\frac{\lambda_{p}^{yy}}{\lambda_{n}}\neq -2$. 

Furthermore, we calculate $\lambda_{n}$, $\lambda_{\Delta P}$ 
and $\frac{\lambda_{p}^{yy}}{\lambda_{n}}$ 
up to $\delta^{2}$ and $\epsilon^{6}$.
Tables \ref{Table 4.} and \ref{Table 5.} give the values of these constants 
for $\epsilon=0.05$ and 0.1, respectively. 
 The value and sign of $\Delta P$ depends on
$\epsilon$ and ${\rm K\!n}$, too. 
\begin{table}[htb]
\begin{center}
\begin{tabular}{c||c|c|c}
$\rm K\!n$ & $\lambda_{n}\times 10^{-2}$ & $\lambda_{\Delta P}\times 10^{-2}$ & $\lambda_{p}^{yy}/\lambda_{n}$ \\
\hline \hline
0.005 & -0.552682 & 1.59241 & -2.72540 \\ 
\hline
0.01  & 0.3430882 &-0.16459 & -1.00243 \\
\hline
0.02  & 0.2577696 &-0.22015 & -1.63486 \\
\hline
0.05  & 0.1182556 &-0.11280 & -1.87030 \\
\hline
0.1   & 0.0612808 &-0.05996 & -1.93731 \\
\hline
0.2   & 0.0311473 &-0.03082 & -1.96915 \\
\hline
\end{tabular}
\end{center}
\caption{The values of 
$\lambda_{n}$, $\lambda_{\Delta P}$ and $\lambda_{p}^{yy}/\lambda_{n}$ 
for $\epsilon=0.05$ in the case of EG without a strong nonlinear term.}
\label{Table 4.}
\end{table}
\newpage

\begin{table}[ht]
\begin{center}
\begin{tabular}{c||c|c|c}
$\rm K\!n$ & $\lambda_{n}\times 10^{-2}$ &$\lambda_{\Delta P}\times 10^{-2}$ & 
$\lambda_{p}^{yy}/\lambda_{n}$ \\
\hline \hline
0.005 &-0.598881 & 1.66316 & -2.60733 \\ 
\hline
0.01  & 0.340322 &-0.15760 & -0.98090 \\
\hline
0.02  & 0.259368 &-0.22086 & -1.62996 \\
\hline
0.05  & 0.119439 &-0.11385 & -1.86873 \\
\hline
0.1   & 0.061951 &-0.06598 & -1.93656 \\
\hline
0.2   & 0.031501 &-0.03117 & -1.96879 \\
\hline
\end{tabular}
\end{center}
\caption{The values of 
$\lambda_{n}$, $\lambda_{\Delta P}$ and $\lambda_{p}^{yy}/\lambda_{n}$ 
for $\epsilon=0.1$ in the case of EG without a strong nonlinear term.}
\label{Table 5.}
\end{table}

%%%%%%%%%%%%%%%%%%%%%%%%%%%%%%%%%%%%%%%%%%%%%%%%%%%%%%%%%%%%%%%%%%%%

\section{Discussion}\label{sec:discussion}

In Ref.~\cite{sasa} the authors argue that
there is a  pressure difference at plate 2, namely the pressure in one side
of the plate is different to that on the other side. 
They call this new pressure which acts on the central plate the ``flux induced
osmosis'' (FIO). 
We consider the existence of FIO proposed by ~\cite{sasa} 
identifying $\Delta P$ as the pressure difference defined in Sec. 
\ref{sec:consideration}.

In Ref.~\cite{sasa} the following criteria are stated:

1. $\Delta P>0$ regardless of the sign of $Q_{y}$. 

2. $P_{yy}$ is a function of the nonequilibrium quantities:  
$T_{1}$,  the nonequilibrium steady heat flow $Q_{y}$, 
and it is related to the equilibrium quantity $P_{\sf eq}$ 
as long as the nonequilibrium 
and equilibrium temperature at both sides of plate 2    
coincide, 
\begin{equation}\label{sstb} 
\frac{n(0)}{n_{\sf  eq}} = \left(\frac{\partial
P_{yy}}{\partial P_{\sf eq}}\right)_{T_{1},Q_{y}} 
\end{equation} 
where $T_{1}$ is the thermodynamic temperature of plate 2.

In Ref.~\cite{kim}, for a system of  hard spheres 
and of  maxwellian particles---which obey  Boltzmann's equation 
and which obey the BGK equation~\cite{BGK}---using the Chapman-Enskog method,
it is shown that criterion 1 in Ref.~\cite{sasa} is valid 
but criterion 2 is not valid. 

On the other hand, for the hard disk's system, from our present
scheme based on Enskog's equation we obtain that

1. Criterion 1 in Ref.~\cite{sasa} is not obeyed in the case
of NLBG: $\Delta P$ is always {\em negative} independent of the sign of $Q_{y}$.
 
2. In the case of LBG, (\ref{sstb}) is valid. 
However, substitution of (\ref{lambdan}) and (\ref{lambdapyy}) 
into (\ref{sstb}) leads to $\lambda_{p}^{yy}/\lambda_{n}=-2$. 
This is correct  only in the case of NLBG. 
Hence, criterion 2 in Ref.~\cite{sasa} is not satisfied either.

In the formulation of SST  it is assumed that the number of
particles and the size of the system is infinite but that the
number density of the system is finite~\cite{sasa}. This condition implies
that the terms  $O(\delta^2)$ in the collision terms can be neglected.
Table \ref{Table 3.} still indicates that under such condition the osmotic
pressure difference is not always positive. Especially when the system is
extremely close to equilibrium, Eq.~(\ref{leg1}) implies that $\Delta P$ is
always negative. 

However when the system is extremely close to equilibrium, 
Eq.~(\ref{leg2}) implies that $\Delta P$ is
always positive.  This result coincides with the hard sphere
case in Ref.~\cite{kim}.

\bigskip             

Furthermore, the condition under which there is no heat flux  
at the porous wall is 
\begin{equation}\label{tsuriH}
\int_{-\infty}^{\infty}dc_{x}\int_{0}^{\infty}dc_{y}
                     \frac{c_{y}}{2}C^{2}f_{\rm equil.} 
        +\int_{-\infty}^{\infty}dc_{x}\int_{-\infty}^{0}dc_{y}
                     \frac{c_{y}}{2}C^{2}f_{y=0}=0. 
\end{equation} 
Using Eq. (\ref{tsuri}) yields
\begin{equation}\label{keyosH} 
q_{y}^{k}(0) = \sqrt{\frac{2}{\pi}}[n_{\sf  eq}-n(0)].                           
\end{equation} 
The above condition is not satisfied without introducing a difference
between the temperature of the gas and the porous wall 
except in the LBG case.  
Therefore, except in the LBG case, it must be difficult 
to maintain the equilibrium state between plates 1 and 2 
even if the heat conductivity of plate 2 is extremely high. 
Here, as a rough simplification,  let us introduce the temperature of the gas 
in contact with plate 2,  $T_{g}\neq T_{1}$. 
Substituting  $\delta=0$  
in Eqs.(\ref{keyos}),  (\ref{EGeq1}), (\ref{EGeq2}), (\ref{EGeq3}), 
(\ref{peq}), (\ref{osmop}), (\ref{lambdan}), (\ref{lambdadp}) and
(\ref{lambdapyy}),  using the  no-mass-flux condition given 
by an equation similar to (\ref{tsuri}), yields, 
\begin{equation}\label{jump}
\Delta P=n(0)\left[1-\frac{\sqrt{T_{g}}}{2}-\frac{1}{2\sqrt{T_{g}}}\right]
+\left[\frac{1}{\sqrt{T_{g}}}-2\right]\frac{Q_{y}^2}{256n(0)}.
\end{equation}   
{\em As $T_{1}<T_{g}<T_{2}$ it follows that $1<T_{g}<1+\epsilon$, 
and putting $q_{y}^{k}(y)^{2}=Q_{y}^2=0$ 
in the right side of Eq. (\ref{EGeq3}),
not only for NLBG but also for LBG, $\Delta P<0$.}
The behavior of the pressure difference changes qualitatively 
even if we restrict the analysis to Boltzmann's regime.
This implies that the estimation of $\Delta P$ is a very delicate problem.
Even if one can prepare the walls which satisfy Eq.~(\ref{sstb}) 
and estimate the pressure difference in  Enskog's regime,
it is difficult to know what  physical meaning 
lies behind such case.

\bigskip

%%%%%%%%%%%%%%%%%%%%%%%%%%%%%%%%%%%%%%%%%%%%%%%%%%%%%%%%%%%%%%
%%%%%%%%%%%%%%%%%%%%%%%%%%%%%%%%%%%%%%%%%%%%%%%%%%%%%%%%%%%%
\section{Conclusions} \label{sec:conclusion}

We have analyzed a simple nonequilibrium steady state system inspired
by~\cite{sasa}. Our study refers only to a hard disk system and analyze in
great detail its behavior using our extended hydrodynamic
equations~\cite{hu} using various approximations.   Since we obtain that 
the osmotic pressure difference  is negative in 
many cases for which  Eq.~(\ref{sstb})
is not satisfied  we cannot agree with~\cite{sasa}. 

%%%%%%%%%%%%%%%%%%%%%%%%%%%%%%%%%%%%%%%%%%%%%%%%%%%%%%%%%%%%%%%%%%%

We have assumed that the pores in plate 2 are small enough 
and we have not considered the problem about reflections on the wall at all. 
As we point out in the last part of Sec. \ref{sec:discussion},
the boundary (contact) condition is very delicate.
We recognize that a more  sophisticated analysis is necessary.

However, in cases when strong nonlinearities can be neglected and  the
system is quite close to equilibrium---so that higher order terms in
$\epsilon$ do not contribute---then the osmotic pressure is positive.
This implies the possibility of the existence of FIO.

In addition, in the full-paper~\cite{sasa}, the authors
point out that 
Eq. (\ref{sstb}) is directly related to the condition at the wall and this
condition is essential to construct the formalism of SST in a complete form
which gets a new nonequilibrium extensive quantity which determines the
degree of nonequilibrium.  Therefore, to clarify the problem, the measure of
the pressure difference is done only for the case of a wall obeying Eq.
(\ref{sstb}).  Within this context, they still recognize the results
of~\cite{kim} as implying the existence FIO.  Hence we also think that it is
worth estimating the pressure difference starting from Eq. (\ref{sstb}). If
one were to analyze the problem in such a way then the boundary (contact)
condition would have to be reconsidered to solve the kinetic equation. In
other words, one would have to evaluate $\Delta P$ under the rather complex
conditions required by kinetic theory that would lead to satisfy
Eq.~(\ref{sstb}). This has not been done.

The SST formalism is quite interesting and the present study has only put to
test  the possible existence of FIO.

Finally, we briefly comment about EIT:
Extended irreversible thermodynamics~\cite{EIT,dominguez,camacho}.
For an ideal gas, Refs.~\cite{dominguez,camacho} studied a
problem quite similar to the one in the present article.
In Ref.~\cite{dominguez} the authors estimated the pressure difference 
without considering a special wall. 
They assumed that the direction of the 
heat flow is parallel to the interface and their results 
are very interesting.  
Furthermore, in~\cite{camacho} the authors estimated the difference 
between the pressures which are parallel and perpendicular to the heat flow.

For a hard disk, appling LBG to the systems of ~\cite{dominguez,camacho}, 
it is easily  possible to get similar equations to (\ref{LBGeq}) 
and to see that the pressure difference
predicted by ~\cite{dominguez} is positive and that the difference 
predicted by ~\cite{camacho} is zero. In both cases, of course, 
 it is totally unnecessary to use  conditions (\ref{tsuri}) 
and (\ref{masssum}).

%%%%%%%%%%%%%%%%%%%%%%%%%%%%%%%%%%%%%%%%%%%%%%%%%%%%%%%%%%%%%%%%%%%%
%%%%%%%%%%%%%%%%%%%%%%%%%%%%%%%%%%%%%%%%%%%%%%%%%%%%%%%%%%%%%%%%%%%
\bigskip
 
{\bf Acknowledgements}

{The author wishes to express his sincere gratitude to Prof. P.
Cordero who
has guided him into the investigation using kinetic theory, given him the
basic theme related to the present investigation and conducted many
important discussions.  The author also wishes to thank Prof. Dino Risso and
Prof. Rodrigo Soto for many helpfull hints.  The scholarship from {\em
Mecesup UCh 0008 project} is gratefully acknowledged.}

%%%%%%%%%%%%%%%%%%%%%%%%%%%%%%%%%%%%%%%%%%%%%%%%%%%%%%%%%%%%%%%%%

\end{document}